\documentclass{elsart}

\usepackage{amssymb}
\usepackage{epsfig}

\begin{document}

\begin{frontmatter}

\title{Relevant distance between two different instances of the same
potential energy in protein folding}

\author[BIFI,FT]{Jos\'e L. Alonso}
\author[BIFI,FT]{Pablo Echenique\corauthref{correspondence}}
\corauth[correspondence]{Correspondent author. {\it Email address:} 
{\tt pnique@unizar.es}}
\address[BIFI]{Instituto de Biocomputaci\'on y F\'{\i}sica de los Sistemas
Complejos (BIFI), Edificio Cervantes, Corona de Arag\'on 42, 50009, Zaragoza,
Spain.}
\address[FT]{Departamento de F\'{\i}sica Te\'orica, Facultad de Ciencias,
Universidad de Zaragoza, Pedro Cerbuna 12, 50009, Zaragoza, Spain.}

\begin{abstract}
In the context of complex systems and, particularly, of protein
folding, a physically meaningful distance is defined which allows to
make useful statistical statements about the way in which energy
differences are modified when two different instances of the same
potential-energy function are used. When the two instances arise from
the fact that different algorithms or different approximations are
used, the distance herein defined may be used to evaluate the relative
accuracy of the two methods. When the difference is due to a change in
the free parameters of which the potential depends on, the distance
can be used to quantify, in each region of parameter space, the
robustness of the modeling to such a change and this, in turn, may be
used to assess the significance of a parameters' fit.  Both cases are
illustrated with a practical example: the study of the
\mbox{Poisson--based} solvation energy in the \mbox{Trp--Cage} protein
(PDB code 1L2Y).
\end{abstract}


\end{frontmatter}

\section{Introduction}
\label{sec:introduction}

The most fundamental way to account for the behaviour of a physical
system is through its energy function $\mathcal{H}({\vec q},{\vec
p})$, which depends on the coordinates $\vec q$ and the momenta $\vec
p$ of all the particles.  In normal situations, this function can be
expressed as the sum of the kinetic energy $\mathcal{K}({\vec q},{\vec
p})$\footnote{The kinetic energy $\mathcal{K}$ depends, in general, on
the positions and the momenta.  However, if cartesian coordinates are
used, the dependence on positions vanishes.} and the potential energy
$\mathcal{V}({\vec q})$.  Since the former is of general form for any
type of system and, normally, it does not affect the equilibrium
properties, the latter is enough for a complete characterization of
the problem.

Under most real circumstances, the exact form of $\mathcal{V}({\vec
q})$ is unknown and one is forced to seek an approximation
$\mathcal{V}^{\mathrm{app}}({\vec q})$. This may be done, for
physical systems that are significantly complex, by assuming that the
relevant interactions included in $\mathcal{V}({\vec q})$ can be
formally factorized \cite{PE:Dil1997JBC}. Then, an approximated function
$\mathcal{V}^{\mathrm{app}}_{i}({\vec q})$ is devised according to
heuristic and semiempirical reasons to account for each of the
original parts $\mathcal{V}_{i}({\vec q})$:

\begin{equation}
\label{eq:fact_ex}
\mathcal{V}({\vec q})=\sum^{m}_{i=1}\mathcal{V}_{i}({\vec q}) \simeq
  \sum^{m}_{i=1}\mathcal{V}^{\mathrm{app}}_{i}({\vec q}) \, .
\end{equation}

For example, in the study of proteins
\cite{PE:Dil1999PS,PE:Alo2004BOOK}, which are a very relevant case of
complex systems, some of the terms in which the total potential-energy
function is traditionally factorized are the hydrogen-bonds energy,
the Van der Waals interaction, the excluded-volume repulsion, the
Coulomb energy and the solvation energy.  This last interaction, which
is one of the most challenging terms of $\mathcal{V}({\vec q})$ to
model, is customarily further split into electrostatic and
non-electrostatic parts \cite{PE:Laz2003BC,PE:Rou2001BOOK}. It is the
former which is studied in section~\ref{sec:application} to illustrate
the application of the concepts herein discussed.

Returning to the general case, let us assume that a particular energy
term $\mathcal{V}_{j}({\vec q})$ in eq.~\ref{eq:fact_ex} and its
approximated counterpart $\mathcal{V}^{\mathrm{app}}_{j}({\vec q})$
correspond to the part of $\mathcal{V}({\vec q})$ that is going to be
studied or modeled.  Let us denote that term $\mathcal{V}_{j}({\vec
q})$ by $V({\vec q})$ and, correspondly,
$\mathcal{V}^{\mathrm{app}}_{j}({\vec q})$ by $V^{\mathrm{app}}({\vec
q})$ in the forecoming reasoning.  Clearly, if the approximated
function $V^{\mathrm{app}}({\vec q})$ is {\it too distant} from the
original $V({\vec q})$, it will be useless, as this difference will
certainly propagate to the total potential energy.  Therefore, one
must precisely define and calculate this {\it distance}, depending on
the type of system which is the object of the study and on the
particular aspects that are going to be investigated.  The situation
is further complicated when subsequent approximations to
$V^{\mathrm{app}}({\vec q})$ are done, usually with the aim of
lessening the numerical complexity and rendering the simulations
feasible.

This yet-undefined distance between potential-energy functions may
also be useful in another situation which is often found in the study
of complex systems: parameter fitting.  Any reasonable functional form
of a certain term $V({\vec q})$ of the total potential energy (or of
its approximation $V^{\mathrm{app}}({\vec q})$) is a simplified model
of physical reality and it contains a number of free parameters.
These parameters, which, in most of the cases, are not physically
observable, must be fit against experimental or more {\it ab initio}
results prior to using the function for practical purposes.  For
example, in the continuum solvation models based on the solution of
the Poisson equation
\cite{PE:Rou2001BOOK,PE:Hon1995SCI,PE:Oro2000CR,PE:Zha1997JFI},
typical free parameters are the dielectric constants and the position
of the surface that separates the outer \mbox{high--dielectric} medium
from the inner \mbox{low--dielectric} one.  Although they are
customarily assigned {\it standard} values (such as ${\kappa}_{P}=1$
for the dielectric constant of the protein, ${\kappa}_{W}=80$ for the
dielectric constant of the aqueous medium\footnote{
\label{foot:kappa} In the field of
molecular simulations, $\kappa$ usually denotes the
\mbox{Debye-H\"uckel} parameter, which quantifies the ionic strength
in the aqueous medium; $\epsilon$ is customarily used to represent the
dielectric constant. However, in this work, the usual convention in
physics, by which $\kappa$ stands for the dielectric constant and
$\epsilon$ stands for the dielectric permittivity, is preferred. Since
the ionic strength is zero in all calculations, this choice should not
be misleading.}, and the Molecular Surface (MS), defined by Connolly
\cite{PE:Con1983JAC}, for the surface of separation), we believe that
they must be fit in order to render calculations more accurate.
Certainly, for other potential-energy functions, such as the ones
found in force fields like CHARMM \cite{PE:Bro1983JCC,PE:Mac1998BOOK},
the fit of the free parameters is common practice.

In order for any fit to yield statistically significant values of the
parameters, the particular region of the parameter space in which the
final result lies must have the property of {\it robustness}, i.e., it
must occur that, if the found set of parameters' values is slightly
changed, then, the relevant characteristics of the potential-energy
function which depends on them are also approximately kept unmodified.
If this were not the case, a new fit, performed using a different set
of experimental (or more {\it ab initio}) points, could produce a very
{\it distant} potential.  This last scenario is, clearly,
undesirable. Therefore, it may be useful to evaluate the robustness of
the, {\it a priori}, reasonable regions of parameter space for the
potential energy function that is to be used.  To accomplish this, one
must again define a relevant distance between two {\it instances} of
the same potential with different values of the parameters.

In section~\ref{sec:distance}, a meaningful distance that can be used
in the two situations aforementioned is defined and justified.  In
section~\ref{sec:application}, within the context of the protein
folding problem and as an example of the first application discussed,
this distance is measured between instances of the
\mbox{Poisson--based} solvation energy arising from the choice of
different grid sizes in the finite-differences algorithm with which it
is calculated.  To illustrate the second possible use of the distance,
the robustness of the Poisson energies to changes in some of its free
parameters, holding the grid size fixed, is quantified.  This analysis
is necessary to assess the significance of the parameters' values
determined through a fit.  Finally, section~\ref{sec:conclusions} is
devoted to the conclusions.

\section{Distance criterium}
\label{sec:distance}

Let $V({\vec q})$ be a particular term of the potential energy of a
system. The numerical value of this physical quantity for each
conformation ${\vec q}$ depends on two conceptually different inputs:
on one hand, the algorithm or approximation used to compute it,
denoted by $\mathcal{A}$; on the other hand, the values of the free
parameters ${\vec P}$.  Changes in these inputs produce different {\it
instances} of the physical quantity $V({\vec q})$, which we denote by
subscripting $V$.  For example, if the algorithm is held constant and
two different set of parameters ${\vec P}_{1}$ and ${\vec P}_{2}$ are
used, our notation made explicit would read as in the following
equation (an analogous definition may be made if the algorithm is held
constant and the parameters are varied).

\begin{equation}
\label{eq:notation}
V_{1}({\vec q}) := V(\mathcal{A},{\vec P}_{1},{\vec q}) \qquad
\mathrm{and} \qquad
V_{2}({\vec q}) := V(\mathcal{A},{\vec P}_{2},{\vec q}) \, .
\end{equation}

Now, a useful and physically meaningful definition of a distance
$d(V_{1},V_{2})$ is sought between a pair of instances, such as the
ones in the previous equation, of the same potential-energy function.

\begin{figure}[t]
\begin{center}
\epsfig{file=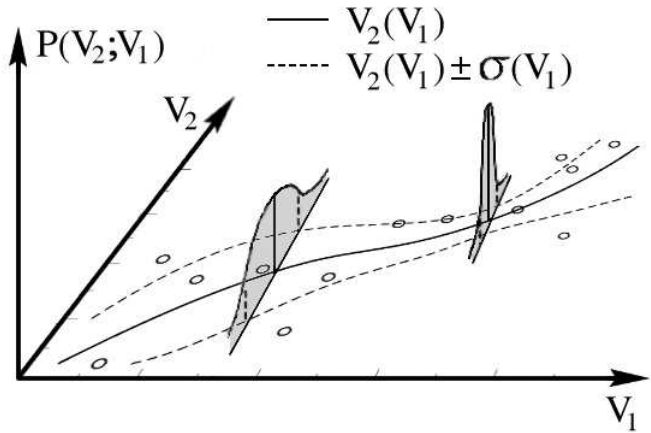,width=9cm}
\end{center}
\caption{{\it Illustration} of the functions defined in
eq.~\ref{eq:mean_std}.  Each point corresponds to a single
conformation ${\vec q}_{i}$ of the system, being $V_{1}({\vec q}_{i})$
its \mbox{x--coordinate} and $V_{2}({\vec q}_{i})$ its
\mbox{y--coordinate}. In the \mbox{z--axis}, the probability density
of {\it measuring} a particular value of $V_{2}$ for each value of
$V_{1}$.  Two normal distributions representing this quantity are
shown at arbitrary positions in the \mbox{x--axis}.  The continuum
line represents the mean $V_{2}(V_{1})$ of the values $V_{2}({\vec
q})$ with ${\vec q} \in \mathcal{C}(V_{1})$ as a function of
$V_{1}$. The broken lines enclose the region where there is the
largest probability to find a conformation if a single {\it numerical
experiment} is performed (around $68\%$ if the distribution of $V_{2}$
is assumed to be normal for each $V_{1}$).}
\label{fig:ideal_function_3D}
\end{figure}

In some cases traditionally studied in physics, the dependence of $V$
on the parameters is simple enough to allow a closed functional
dependence $V_{2}(V_{1})$ to be found\footnote{
\label{foot:simple_P}
For example, if the mass of a harmonic oscillator is changed from
$m_{1}$ to $m_{2}$, the potential energy functions will satisfy the
linear relation \mbox{$V_{2}({\vec q})=(m_{1}/m_{2})V_{1}({\vec q})$}
for all the conformations of the system; if the the atomic charges are
rescaled by a factor $\alpha$ (being actually ${\alpha}Q_{i}$) and
$\alpha$ is changed from ${\alpha}_{1}$ to ${\alpha}_{2}$, the free
energies of solvation calculated via the Poisson equation will, in
turn, satisfy the linear relation \mbox{$V_{2}({\vec
q})=({\alpha}_{1}/{\alpha}_{2})^{2}V_{1}({\vec q})$}, etc.}.  However,
in the study of complex systems, such as proteins, this dependence is
often much more complicated, due to the high dimensionality of the
conformational space and to the fact that the energy landscape lacks
any evident symmetry.  The set $\mathcal{C}(V_{1})$ of the
conformations with a particular value of the potential energy $V_{1}$
typically spans large regions of the phase space containing
structurally different conformations.  When an approximation is
performed or the algorithm is changed, from $\mathcal{A}_{1}$ to
$\mathcal{A}_{2}$, or the free parameters are shifted, from ${\vec
P}_{1}$ to ${\vec P}_{2}$, each conformation ${\vec q}$ in
$\mathcal{C}(V_{1})$ is affected in a different way and its new energy
$V_{2}({\vec q})$ is modified in a manner that does not depend trivially
on the particular region of the phase space which the conformation
${\vec q}$ belongs to.  Therefore, a simple functional relation
$V_{2}(V_{1})$ is no longer possible to be found: for each value of
$V_{1}$, there corresponds now a whole distribution of values of
$V_{2}$ associated with conformations which share the same value of
$V_{1}$ but which are far apart in the conformational space.
Moreover, the projection of this \mbox{high--dimensional} \mbox{${\vec
q}$--space} into the \mbox{1--dimensional} \mbox{$V_{1}$--space makes}
$V_{2}$ look as a random variable for each particular value of $V_{1}$
(see fig.~\ref{fig:ideal_function_3D}).  We then define two real
functions, $V_{2}(V_{1})$ and ${\sigma}(V_{1})$, which correspond to
the mean and to the standard deviation, respectively, of this random
variable as a function of $V_{1}$ (where the average $\langle \cdot
\rangle$ is defined to be taken over the conformations ${\vec q}
\in \mathcal{C}(V_{1})$):

\begin{equation}
\label{eq:mean_std}
V_{2}(V_{1}) := \langle V_{2}({\vec q}) \rangle \qquad
\mathrm{and} \qquad
{\sigma}(V_{1}) := \sqrt{\langle \,
  (V_{2}({\vec q})-\langle V_{2}({\vec q}) \rangle)^{2} \, \rangle}
  \, .
\end{equation}

It can be proved, from their definition, that these two functions are
continuous irrespectively of the particular characteristics of
$V_{1}({\vec q})$ and $V_{2}({\vec q})$ (given that both of them are
{\it smooth} functions of the conformation). It may also be shown
that, under assumptions which are typically fulfilled in real cases,
$V_{2}(V_{1})$ and ${\sigma}(V_{1})$ are also differentiable.  Thus,
when restricted to a finite interval of $V_{1}$, the linear dependence
given by the following equation may hold approximately:

\begin{equation}
\label{eq:linear}
V_{2}(V_{1}) \simeq bV_{1}+a:=b(V_{1}+V_{1}^{0}) \, .
\end{equation}

In fact, for the aforementioned cases in which the dependence of the
potential energy on the parameters is simple enough (see footnote
\ref{foot:simple_P}), eq.~\ref{eq:linear} holds exactly and, being
able to describe $V_{2}(V_{1})$ by a closed analytical formula, $b$
and $a$ can be exactly computed\footnote{ One must be careful about
the notation. Although the function $V_{2}(V_{1})$ has been formally
defined in eq.~\ref{eq:mean_std} as an average, in the case of the
simple systems in footnote~\ref{foot:simple_P}, a function
$V_{2}(V_{1})$ may be found algebraically. This is, however,
consistent.  If one thinks that, in these systems, the spread
${\sigma}(V_{1})$ is zero, the relation found algebraically and the
one that stems from the average are identical.}.  However, in a
general situation, the functions defined in eq.~\ref{eq:mean_std} are
impossible to be calculated analytically and so are the parameters $b$
and $a$ in the linear approximation of $V_{2}(V_{1})$.  In such a
case, one may at most have a finite collection of $n$ conformations
$\{{\vec q}_{i}\}^{n}_{i=1}$ and the respective values of $V_{1}({\vec
q}_{i})$ and $V_{2}({\vec q}_{i})$ for each one of them.  These data,
for a particular $i$, should be thought as a single {\it numerical
experiment} in the already suggested sense that, if one regards
$V_{1}({\vec q}_{i})$ as an independent variable, the {\it outcome} of
the dependent variable $V_{2}({\vec q}_{i})$ is basically
unpredictable and $V_{2}$ may be regarded as a random variable
parametrically dependent on $V_{1}$ (see
fig.~\ref{fig:ideal_function_3D}).  From this finite knowledge about
the system, one may statistically estimate the values of $b$ and
$a$. If the standard deviation ${\sigma}(V_{1})$ is a constant (i.e.,
it does not depend on $V_{1}$), which, for the particular system
studied in this work, has been found to be approximately true, then,
the least-squares maximum-likelihood method
\cite{PE:Pre2002BOOK,PE:Bev2003BOOK} yields the best estimates for $b$
and $a$\footnote{
\label{foot:estimates}
The same letters are used for the ideal parameters $b$, $a$, and
$\sigma$ and for their least-squares best estimates, since the only
knowledge that one may have about the former comes from the
calculation of the latter.} under very general conditions, such as
independence of the random variables and normal distribution.
Moreover, it may be shown \cite{PE:Pre2002BOOK,PE:Bev2003BOOK} that
the best estimate for the standard deviation $\sigma$ (see
footnote~\ref{foot:estimates}) is given by the following expresion:

\begin{equation}
\label{eq:sigma}
\sigma \simeq \sqrt{\frac{{\sum}_{i=1}^{n}
    [V_{2}({\vec q}_{i})-(bV_{1}({\vec q}_{i})+a)]^{2}}{n-2}} \, .
\end{equation}

This quantity $\sigma$ may be regarded as a {\it random error} that
arises in the transit from $V_{1}$ to $V_{2}$. In the same sense, $a$
may be regarded as a {\it systematic error} and, since it is
equivalent to an energy reference, its actual value is not relevant
for the forecoming discussion. The slope $b$ and $\sigma$ are the two
quantities involved in the definition of the distance
$d(V_{1},V_{2})$, which is the central concept introduced in this
paper (see eq.~\ref{eq:def_d} below). In order to render this
definition meaningful, we are going to evaluate how {\it energy
differences} are modified, when going from $V_{1}$ to $V_{2}$, as a
function of $b$ and $\sigma$. This differences are the relevant
physical quantities if one's aim is to study the conformational
behaviour of a system.

\begin{figure}[t]
\begin{center}
\epsfig{file=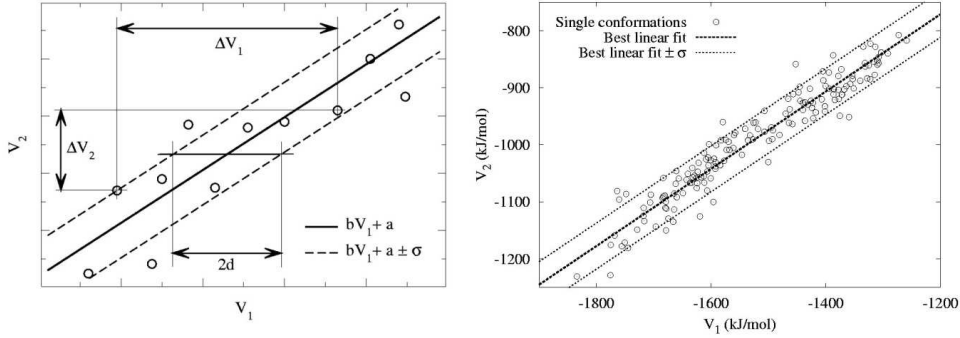,width=13cm}
\end{center}
\caption{The graphic on the left is an {\it illustration} of the
distance criterium. The best linear fit is depicted by a continuum
line. The broken lines correspond to the estimated standard deviation
of the points in the $V_{2}$ direction. The $V_{1}$-- and
$V_{2}$--energy differences, ${\Delta}V_{1}$ and ${\Delta}V_{2}$,
between two particular conformations, as well as the distance $d$ for
${\alpha}=1$ (see eq.~\ref{eq:def_d}) are also shown. The graphic on
the right is a particular example taken from the system studied in
section~\ref{sec:application}. The difference between the $V_{1}$ and
$V_{2}$ instances is a small modification of the surface of separation
between dielectrics. 150 conformations of the investigated protein are
shown.}
\label{fig:criterium}
\end{figure}

Under the approximations of linear $V_{2}(V_{1})$ dependence and
constancy of ${\sigma}(V_{1})$, fig.~\ref{fig:ideal_function_3D}
becomes the left part of fig.~\ref{fig:criterium}.  In the right side
of the same figure, one of the {\it worst} cases (i.e., one case for
which $d(V_{1},V_{2})$ is one of the largest) studied in
section~\ref{sec:application} is depicted to show that the hypothesis
are approximately fulfilled for the particular protein system
investigated there.

Now, let us focus on two arbitrary conformations of the system (see
the left side of fig.~\ref{fig:criterium}). For them, the
\mbox{$V_{2}$--energy} difference ${\Delta}V_{2}$ is a random variable
parametrically dependent on the \mbox{$V_{1}$--energy} difference
${\Delta}V_{1}$ (which is assumed to be regular number, i.e., a random
variable with zero variance\footnote{See the discussion near the end
of the section for a more detailed analysis of the implications of
this assumption.}). The probability density of this random variable is
found by assuming that $V_{2}$ is normally distributed with mean
$V_{2}(V_{1})$ and standard deviation $\sigma$. Then, the distribution
of ${\Delta}V_{2}$, for each ${\Delta}V_{1}$, is normal with mean
$b{\Delta}V_{1}$ and standard deviation $\sqrt{2}\sigma$\footnote{
\label{foot:statistics}
If $x$, $y$ and $z$ are random variables and the relation $z=Ax+By$
holds, then \mbox{$\langle z \rangle =A\langle x \rangle +B\langle y
\rangle$}.  If $x$ and $y$ are independent, one also has that
\mbox{${\sigma}_{z}^{2}=A{\sigma}_{x}^{2}+B{\sigma}_{y}^{2}$}
\cite{PE:Pre2002BOOK,PE:Bev2003BOOK}.}:

\begin{equation}
\label{eq:pdf_DV2}
\mathcal{P}({\Delta}V_{2};{\Delta}V_{1})=
    \frac{1}{\sqrt{2\pi}(\sqrt{2}\sigma)}
    \exp \left [ -\frac{({\Delta}V_{2}-b{\Delta}V_{1})^{2}}
	 {2(\sqrt{2}\sigma)^{2}} \right ] \, .
\end{equation}

If the random errors were negligible (as in the systems discussed in
footnote~\ref{foot:simple_P}) and one wanted to calculate the value of
${\Delta}V_{1}$ from the knowledge of ${\Delta}V_{2}$, the identity
\mbox{${\Delta}V_{1}={\Delta}V_{2}/b$} would have to be used. When
there are significant random errors, the situation is equivalent
except for the fact that there is a probabilistic indetermination,
i.e., from the {\it measured} value of ${\Delta}V_{2}$,
${\Delta}V_{1}$ can be no longer inferred. It follows directly from
eq.~\ref{eq:pdf_DV2} that, the quantity ${\Delta}V_{2}/b$ is a random
variable normally distributed with mean ${\Delta}V_{1}$ and standard
deviation $\sqrt{2}\sigma /|b|$ (see footnote~\ref{foot:statistics}):

\begin{equation}
\label{eq:pdf_DV2b}
\mathcal{P}({\Delta}V_{2}/b;{\Delta}V_{1})=
    \frac{1}{\sqrt{2\pi}(\sqrt{2}\sigma /|b|)}
    \exp \left [ -\frac{({\Delta}V_{2}/b-{\Delta}V_{1})^{2}}
	 {2(\sqrt{2}\sigma /b)^{2}} \right ] \, .
\end{equation}

Now, let us define the distance $d(V_{1},V_{2})$ between two instances
of the same potential energy as follows\footnote{ Although a negative
value of $b$ may look physically perverse, there is no theoretical
drawback about it and the possibility is allowed.  In fact, if the
random errors were small and $b$ was not very small in absolute value,
the loss of information (see the forecoming discussion) in going from
$V_{1}$ to $V_{2}$ would be small.}:

\begin{equation}
\label{eq:def_d}
d(V_{1},V_{2}):={\alpha}\frac{\sigma}{|b|}=\frac{\alpha}{|b|}
    \sqrt{\frac{{\sum}_{i=1}^{n}
    [V_{2}({\vec q}_{i})-(bV_{1}({\vec q}_{i})+a)]^{2}}{n-2}} \, .
\end{equation}

Where $b$ and $a$ are those calculated from a least-squares fit of the
values of $V_{2}({\vec q}_{i})$ against $V_{1}({\vec q}_{i})$ and
$\alpha$ is a positive proportionality factor yet to be fixed (see
point 3 below).

This definition encodes some intuitions that one may have about the
{\it loss of information} involved in the transit from $V_{1}$ to
$V_{2}$.  Let us remark some important properties which illustrate
this fact:

\begin{enumerate}
\item If the slope $b$ is nonzero and the random error $\sigma$ is
 zero, \mbox{$d(V_{1},V_{2})=0$} and there is no loss of
 information. An example of this situation is given by the simple
 systems in footnote~\ref{foot:simple_P}: clearly, no loss of
 information may be involved in changing the mass of an harmonic
 oscillator.
\item If the random error $\sigma$ is different from zero and the
 slope $b$ goes to zero, \mbox{$d(V_{1},V_{2}) \rightarrow \infty$}
 and the loss of information is complete.  One may picture this
 situation by making the best-fit line in fig.~\ref{fig:criterium}
 horizontal. In such a case, when two numerical experiments are
 performed, the probability distribution of measuring ${\Delta}V_{2}$
 does not depend on ${\Delta}V_{1}$ (it is normal with zero mean) and
 the information about ${\Delta}V_{1}$ is impossible to recover.
\item For intermediate cases in which both $b$ and $\sigma$ are
 nonzero, all the information about ${\Delta}V_{2}/b$ is found in its
 probability-density function (see eq.~\ref{eq:pdf_DV2b}) and many
 probabilistic statements may be made. For example, it would be
 desirable that the sign of ${\Delta}V_{2}/b$ had a high probability
 of being equal to the sign of ${\Delta}V_{1}$.  This would typically
 keep the correct energy ordering of the conformations when going from
 $V_{1}$ to $V_{2}$.  Making the variable change
 \mbox{$x={\Delta}V_{2}/b-{\Delta}V_{1}$} in eq.~\ref{eq:pdf_DV2b},
 one finds that this probability is given by the following equation
 (assuming, without loss of generality, that
 \mbox{${\Delta}V_{1}>0$}).

\begin{equation}
\label{eq:ordering}
\mathcal{P}_{\mathrm{ordering}}=
    \int_{-{\Delta}V_{1}}^{\infty}\frac{1}{\sqrt{2\pi}(\sqrt{2}d/{\alpha})}
    \exp \left [ -\frac{x^{2}}
	 {2(\sqrt{2}d/{\alpha})^{2}} \right ] \mathrm{d}x \, .
\end{equation}

 For a same value of ${\Delta}V_{1}$, this probability decreases with
 $d$; if $d$ is held constant, it increases with ${\Delta}V_{1}$ (the
 minimum being \mbox{$\mathcal{P}_{\mathrm{ordering}}=1/2$}, either
 for \mbox{${\Delta}V_{1} \rightarrow 0$} or for \mbox{$d \rightarrow
 \infty$}).  If $d$ is small, the probability of the ordering being
 mantained is large, even for pairs of conformations that are close in
 \mbox{$V_{1}$--energy}.  For example, if one takes ${\alpha}=1$, it
 happens that, if \mbox{${\Delta}V_{1}>d$}, then,
 \mbox{$\mathcal{P}_{\mathrm{ordering}}>76\%$}; if
 \mbox{${\Delta}V_{1}>2d$}, then,
 \mbox{$\mathcal{P}_{\mathrm{ordering}}>92\%$}, etc.  Any other choice
 of $\alpha$ would only yield different numerical values for this
 bounds on $\mathcal{P}_{\mathrm{ordering}}$; the qualitative facts
 would be preserved. However, since this values are natural (the
 normal distribution varies rapidly and one would easily get very
 large or very small probabilities), the choice ${\alpha}=1$ is
 considered to be the most convenient herein and it is the one to be
 used in section~\ref{sec:application}.
\item The properties stated in the previous point are direct
 consequences of the more general fact that, as $d$ decreases, so does
 the standard deviation of the random variable ${\Delta}V_{2}/b$
 (which equals $\sqrt{2}d/{\alpha}$) and the distribution becomes
 sharper around the average ${\Delta}V_{1}$ (see
 eq.~\ref{eq:pdf_DV2b}). That is, the smaller the value of $d$, the
 larger the probability of ${\Delta}V_{2}/b$ being close to the {\it
 perfect} value ${\Delta}V_{1}$.
\end{enumerate}

This measure of the distance between two instances of a potential
energy presents some advantages. On one hand, if the approximations on
which it is based (normal distribution of $V_{2}$ for each $V_{1}$,
linear $V_{2}(V_{1})$ dependence, constancy of ${\sigma}(V_{1})$ and
zero variance in the {\it measures} of $V_{1}$) are reasonable, the
statistical statements derived from a particular $d$ value are
meaningful and precise. On the other hand, this statements refer, like
the ones in the points discussed above, to the whole conformational
space. However, we would like to stress that, in this work, we are not
going to give any criterium to decide whether a particular value of
$d$ is sufficiently small for an approximation \mbox{$V_{1}
\rightarrow V_{2}$} to be valid or for the system to be robust to
changes in the free parameters. Such a decission must be taken
depending on the particularities of the system studied (which are
encoded in the total potential energy function $\mathcal{V}({\vec
q})$) and on the questions sought to be answered. Our definition of
$d(V_{1},V_{2})$, being based in characteristics shared by many
complex systems, is of general application.  For example, we are not
going to establish any limit on the accuracy required for a potential
energy function to successfully predict the folding of proteins
\cite{PE:Dil1999PS,PE:Alo2004BOOK}.  We consider this question a
difficult theoretical problem and we believe that it may be possible
{\it a priori} that some special features of the energy landscapes of
proteins (such as \mbox{funnel--like} shape) are the main responsible
of the high efficiency and cooperativity of the folding process
\cite{PE:Dil1999PS,PE:Alo2004BOOK}. If this were the case, a different
procedure for measuring the distance between potential energy
functions could be devised
\cite{PE:Pan1995JCP,PE:Per1996FD,PE:Per1997FD}, as any approximation
which does not significantly alter these special features would be
valid even if the value of $d$ is very large.  However, for the sake
of simplicity, it will be assumed, herein and in
section~\ref{sec:application} that a transit \mbox{$V_{1} \rightarrow
V_{2}$} between instances of the same potential whose $d(V_{1},V_{2})$
value is {\it of the order} of the thermal fluctuations $RT$ is
acceptable\footnote{ In most computational methods and theoretical
descriptions of a system in contact with a thermal reservoir $RT$ is a
relevant energy ($RT$ is preferred to $k_{B}T$ since per mole energy
units are used in this article) and the results will be presented in
units of $RT$.  It appears in the thermodynamical equilibrium
Boltzmann distribution, in which the probability $p_{i}$ of a
conformation ${\vec q}_{i}$ is proportional to $\exp (-V({\vec q}_{i})/RT)$,
it also determines the transition probability $\mathrm{min}[1,\exp
(-(V({\vec q}_{i+1})-V({\vec q}_{i}))/RT)]$ in the Metropolis Monte
Carlo scheme and it is the spread of the stochastic term in the
Langevin equation \cite{PE:Hes2002PHD}.}.

The last point that we must remark in this section is that the
distance introduced in this paper does not satisfy all the properties
that the class of mathematical objects usually referred to as {\it
distances} do satisfy.  For example, the equivalence
\mbox{$\mathcal{D}(x,y)=0 \Leftrightarrow x=y$} becomes, for the
distance in eq.~\ref{eq:def_d}, an implication in only one direction,
i.e., while it is true that \mbox{$V_{1}=V_{2} \Rightarrow
d(V_{1},V_{2})=0$}, the reciprocal is false in general, since, for
example, if \mbox{$V_{2}=BV_{1}+A$}, with $B$ and $A$ non-zero constants, then
$d(V_{1},V_{2})=0$ when, clearly, the two instances are not equal.
Another property of the mathematical distances that is not fulfilled
by $d$ is the one of symmetry, i.e., that
\mbox{$\mathcal{D}(x,y)=\mathcal{D}(y,x)$}. If we denote all the quantities
calculated when going from $V_{1}$ to $V_{2}$ by subscripting them with
the label $12$ and the ones corresponding to the opposite process with $21$,
we have that:

\begin{eqnarray}
\label{eq:d12_d21}
d(V_{1},V_{2}) &=& \alpha
    \sqrt{\frac{{\sum}_{i=1}^{n}
    [V_{2}({\vec q}_{i})/b_{12}-
      (V_{1}({\vec q}_{i})+a_{12}/b_{12})]^{2}}{n-2}} \, ,
 \cr
d(V_{2},V_{1}) &=& \alpha
    \sqrt{\frac{{\sum}_{i=1}^{n}
    [V_{1}({\vec q}_{i})/b_{21}-
      (V_{2}({\vec q}_{i})+a_{21}/b_{21})]^{2}}{n-2}} \, .
\end{eqnarray}

Therefore, if the equality \mbox{$d(V_{1},V_{2})=d(V_{2},V_{1})$} is
to be hold for every set of conformations $\{{\vec
q}_{i}\}^{n}_{i=1}$, one must require that \mbox{$b_{12}=b_{21}=1$}
and \mbox{$a_{12}=-a_{21}$}.  These two relations impose complicated
conditions over the values of $\{V_{1}({\vec q}_{i})\}^{n}_{i=1}$ and
$\{V_{2}({\vec q}_{i})\}^{n}_{i=1}$ which are not generally fulfilled.
The origin of this lack of symmetry is completely consistent with the
assumptions made about the random character of the two instances of
the potential energy to be compared, namely, the hypothesis of zero
variance of $V_{1}$, which places the two potentials on a different
footing.  A more general distance may be defined (J.L. Alonso and
P. Echenique, in preparation) that takes into account a possible
indetermination in the {\it measure} of $V_{1}$ and that places the
two potentials on the same footing. However, some remarks must be made
about this. In the first place, this asymmetry in the role played by
each of the potential-energy instances is totally justified in the
cases for which the situation intended to be modeled is actually
asymmetric; for example, if one's aim is to calculate the distance
between a potential $V_{1}$ and an approximation $V_{2}$, where
$V_{1}$ may be considered either {\it exact} (e.g., quantum
mechanical) or much more accurate than $V_{2}$.  In such a case, the
assumption of zero variance for $V_{1}$ and the difference in the
roles played by both potentials is intrinsic to the situation
studied. As a second remark, it must be stressed that the distance
herein defined was never intended to be a mathematical distance,
although some of the properties demanded to these objects are
satisfactory and fairly intuitive. The meaning of the distance $d$ is
encoded in the statistical statements derived from its value and the
name {\it distance} must be used in a more relaxed manner than the one
traditionally found in mathematics. Finally, it must be pointed out
that there is a situation in which the symmetry of the distance
defined in this work holds, namely, the situation in which
$b:=b_{12}=1$ and $n \rightarrow \infty$. When the number of
conformations $n$ is very large, the statistical estimators $b$ and
$a$ of the slope and the \mbox{y--intercept} of the linear relation
between $V_{1}$ and $V_{2}$ tend to the ideal values (see
footnote~\ref{foot:estimates}) and, in these conditions, the remaining
requirements needed to satisfy symmetry are also fulfilled, i.e., one
has that $b_{21}=1$, $a_{12}=-a_{21}$ and, consequently,
\mbox{$d(V_{1},V_{2})=d(V_{2},V_{1})$}\footnote{ Note that, if one has
$b:=b_{12}=1$ and \mbox{$n \rightarrow \infty$}, the implication
\mbox{$d(V_{1},V_{2})=0 \Rightarrow V_{1}=V_{2}+A$} also hold. Since
$V_{1}$ and $V_{2}$ are physical energies defined up to a reference,
this may be considered the reciprocal of \mbox{$V_{1}=V_{2}
\Rightarrow d(V_{1},V_{2})=0$}.}. This fact is relevant since there
are many situations typically found in which $b:=b_{12}\simeq 1$,
namely, those in which $V_{1}$ and $V_{2}$ are {\it proximate}. This
is the case if one wants to assess the robustness of a
potential-energy function (a slight change in the parameters does not
lead to a completely different energy) or if the approximation
performed is small. The two applications of the distance $d$ to a
particular potential in sec.~\ref{sec:application} are carried out in
cases for which this {\it proximity} is achieved and the symmetry
expected has been checked numerically.

\section{Application}
\label{sec:application}

Most of the finely tuned biomolecular events occur in a complex
environment of unique characteristics: liquid water. Therefore, if
one aims to correctly describe the crucial processes associated with
proteins, DNA and RNA in living beings, a sufficiently accurate
modeling of \mbox{water--water} and \mbox{water--solute} interactions
must be implemented. However, accuracy is not the only criterium to
be followed when designing a solvation model. Numerical complexity of
the methods must be also taken into account, as computational power is
always a limiting resource.  A compromise between these two competing
factors must be reached and precision may be traded for velocity, even
if the understanding of the problem was complete and a great accuracy
could be achieved.

Particularly, in the study of the protein folding problem, the search
for the native state takes place in an astronomically large
conformational space, as early realized by Cyrus Levinthal in 1969
\cite{PE:Lev1969PROC}. Consequently, the internal energy of the
system, which includes the water molecules and the protein, must be
calculated a large number of times and the numerical complexity of the
method chosen to account for the influence of water must be as low as
allowed by the accuracy required to solve the problem.

\begin{figure}[t]
\begin{center}
\epsfig{file=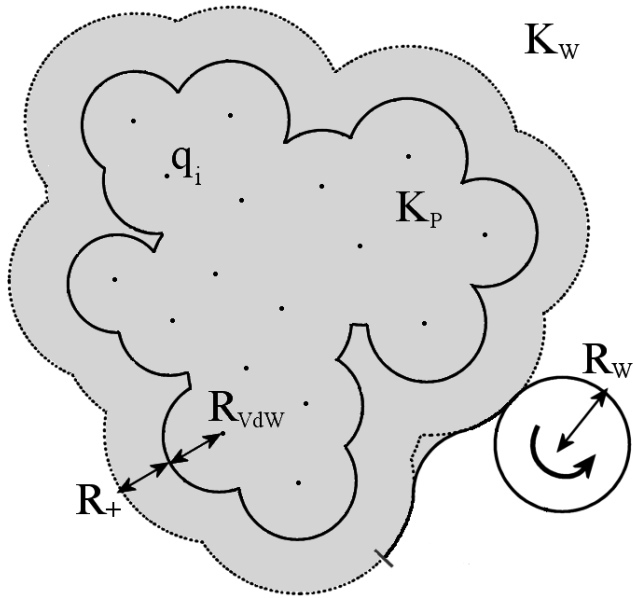,width=9cm}
\end{center}
\caption{Dielectric model of the protein. Atomic charges $q_{i}$ are
punctual and located at nuclei. Space is divided in two disjoint
regions: the macromolecule (in gray), with low dielectric constant
${\kappa}_{P}$ and the water (in white), with high dielectric constant
${\kappa}_{W}$.  The surface of separation between the two media is
constructed as described in the text.}
\label{fig:cavity}
\end{figure}

Despite being regarded as one of the most accurate implementations of
solvent influence, explicit water models are presently too
computationally demanding, allowing only short simulations of peptides
with a small number of residues to be performed. Another popular
option is to use continuum models based on Poisson (PE) or
\mbox{Poisson--Boltzmann} (PBE) equations
\cite{PE:Rou2001BOOK,PE:Hon1995SCI,PE:Oro2000CR,PE:Zha1997JFI}, which
are orders of magnitude faster than explicit solvent models, to
account for the electrostatic part of the free energy of solvation
\cite{PE:Laz2003BC,PE:Rou2001BOOK}.  Then, the nonelectrostatic part,
which arises from the first layer of water molecules surrounding the
solute and from the creation of the cavity, could be added in many
ways, most of which are related to the {\it Solvent Accessible Surface
Area} (SASA) \cite{PE:Eis1986NAT}.

However, it is worth stressing that only the total free energy of
solvation is thermodinamically defined and experimentally
measurable. Consequently, any partitioning of it is necessarily
arbitrary and the free parameters contained in these continuum models
(such as the dielectric constant ${\kappa}_{P}$ of the protein, the
dielectric constant ${\kappa}_{W}$ of the aqueous medium (see
footnote~\ref{foot:kappa}), and the position of the surface that
separates both regions (see fig.~\ref{fig:cavity})), must be fit prior
to use in order to agree with experiment or with more accurate
methods.

In this section, the dielectric constants are set to their standard
values, ${\kappa}_{P}=1$ and ${\kappa}_{W}=80$, and the
characteristics of the surface of separation are modified. Rigorously
speaking, one would need an infinite number of parameters to
completely specify this surface. However, it is used herein a
restricted subset of all the possible surfaces, namely, those that can
be obtained by rolling a sphere of radius $R_{W}$ on the outer side of
the surface generated by adding $R_{+}$ to the Van der Waals
radii\footnote{As found in the {\tt CHARMM23}
\cite{PE:Bro1983JCC,PE:Mac1998BOOK} force field and implemented in the
{\tt pdb2pqr} utility included in the APBS program.} of each atom (see
fig.~\ref{fig:cavity}).  The volume that the rolling sphere does not
intersect is considered to belong to the protein region.  Typical
values assigned to $R_{W}$ and $R_{+}$ in the literature are
\cite{PE:Con1983JAC,PE:Lee1971JMB}:

\begin{enumerate}
\item \mbox{$R_{W}=0.0$ \AA} and \mbox{$R_{+}=0.0$ \AA}, producing the
 {\it Van der Waals Surface}
\item \mbox{$R_{W}=1.4$ \AA} and \mbox{$R_{+}=0.0$ \AA}, producing the
 {\it Molecular Surface}
\item \mbox{$R_{W}=0.0$ \AA} and \mbox{$R_{+}=1.4$ \AA}, producing the
 {\it Solvent Accessible Surface}
\end{enumerate}

These three surfaces are customarily used as the separation between
the two dielectric media when the Poisson energy is
calculated. However, it could be the case that a small change in the
parameters $R_{W}$ and $R_{+}$ significantly alters the properties of
this particular energy landscape.  In such a situation, the choice of
the surface would be crucial to the behaviour of the
system. Therefore, the robustness of the Poisson energy to changes in
$R_{W}$ and $R_{+}$ must be assessed.

\begin{figure}[t]
\begin{center}
\epsfig{file=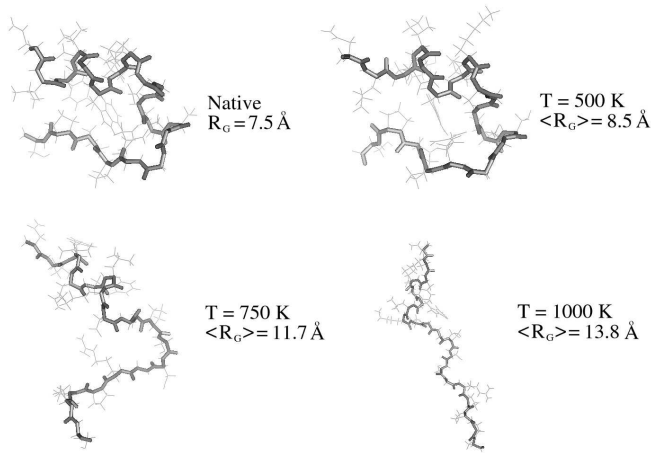,width=9cm}
\end{center}
\caption{Example conformations of the studied \mbox{Trp--Cage}
protein. The native structure, taken from the Protein Data Bank is
shown on the upper left corner. From left to right and from top to
bottom, three particular conformations arbitrarily chosen from three
different sets are depicted in order of decreasing similarity to the
folded protein. The average radii of gyration in each set $\langle
R_{G} \rangle$ and the one of the native structure are also
presented.}
\label{fig:conformations}
\end{figure}

To accomplish this, we study a particular system: the {\it de novo}
designed protein known as \mbox{Trp--Cage} \cite{PE:Nei2002NSB} (PDB
code 1L2Y). The CHARMM molecular dynamics program
\cite{PE:Bro1983JCC,PE:Mac1998BOOK} was used as a conformation
generator. From the native conformation stored in the Protein Data
Bank \cite{PE:Ber2000NAR} a 10 ps heating dynamics\footnote{ The {\tt
c27b4} version of the CHARMM program was used. The molecular dynamics
were performed using the {\it Leap Frog} algorithm therein implemented
and the {\tt param22} parameter set, which is optimized for proteins
and nucleic acids. The water has been taken into account implicitly
with the Dominy {\it et al.} \cite{PE:Dom1999JPC} version of the
Generalized Born Model built into the program.  }, from \mbox{$T=0$ K}
to three different temperatures, \mbox{$T=500$ K}, \mbox{$T=750$ K}
and \mbox{$T=1000$ K}, was performed on the system. This was repeated
50 times for each final temperature with a different seed for the
random numbers generator each time. The overall result of the process
was the production of a set of 150 different conformations of the
protein, 50 of which are {\it close to native}, 50 {\it partially
unfolded} and 50 {\it completely unfolded} (see
fig.~\ref{fig:conformations}). It is worth remarking that the short
time in which the system was heated (10 ps) and the fact that there
was no equilibration after this process cause the final temperatures
(500, 750 and \mbox{1000 K}) to be only {\it labels} for the three
aforementioned sets of conformations. They are, by no means, the
thermodynamical temperatures of any equilibrium state from which the
structures are taken. This three sets of conformations are only meant
to sample the representative regions of the phase space.  In
fig.~\ref{fig:conformations}, one arbitrarily chosen structure from
each set is shown together with the native conformation. The average
radius of gyration $\langle R_{G} \rangle$ of each set, depicted in
the same figure, must be compared to the radius of gyration of the
native state.

Using the finite differences APBS program \cite{PE:Bak2001PNAS}, the
\mbox{Poisson--based} electrostatic part of the solvation energy was
numerically investigated in these conformations. To calculate this
quantity, one must solve the Poisson equation twice. First, the energy
of the system is computed assuming that a dielectric with
${\kappa}={\kappa}_{P}$ fills the whole space. Second, one calculates
the energy of the system with the dielectric geometry shown in
fig.~\ref{fig:cavity}. Finally, the first quantity is substracted from
the second to yield the solvation energy.

\begin{figure}[t]
\begin{center}
\epsfig{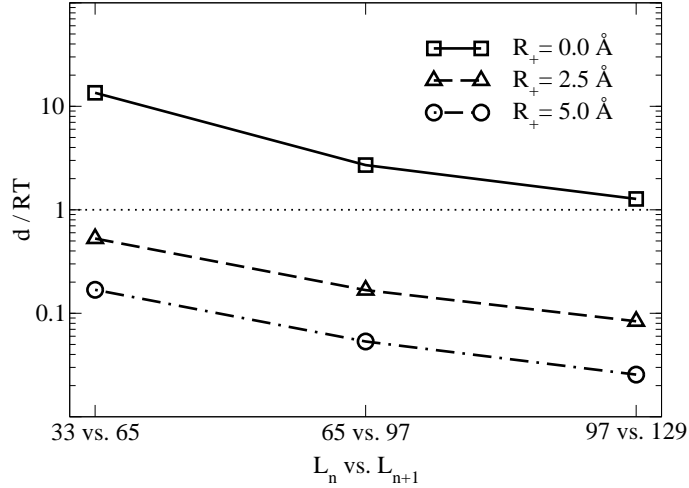}
\end{center}
\caption{Distance between instances of the Poisson solvation energy
with different grid sizes. The \mbox{y--axis} corresponds to the
distance $d$ measured in units of $RT$ (with \mbox{$T=300$ K}) and the
scale is logarithmic. Each point represents the comparison of the
energies calculated with a smaller grid size $L_{n}$ to those
calculated with a larger one $L_{n+1}$. Results for different values
of $R_{+}$ are shown. The value of $R_{W}$ is set to \mbox{1.4 \AA}.}
\label{fig:d_vs_dime}
\end{figure}

In order to test the reliability of the program and as an application
of the first possible use, described in
section~\ref{sec:introduction}, of the distance defined in this paper,
the sensitivity of the Poisson energies to changes in the size of the
grid $L$ used to solve the differential equation was studied. For
algorithmic reasons, the allowed values for $L$ in APBS must be of the
form \mbox{$L_{n}=32n+1$}, with $n$ a positive integer. Consequently,
the Poisson solvation energy of each of the 150 conformations of the
protein was calculated\footnote{ Boundary conditions flag {\tt mdh}
(\mbox{non--interacting} spheres with a point charges), charges' grid
mapping flag {\tt spl2} (cubic \mbox{B--spline} discretization) and
surface smoothing flag {\tt smol} (simple harmonic averaging) were
used in all the calculations.} with \mbox{$L=33$}, \mbox{$L=65$},
\mbox{$L=97$} and \mbox{$L=129$}. All the measures were repeated for
different values of $R_{+}$, (0.0, 2.5 and \mbox{5.0 \AA}) and $R_{W}$
was fixed to \mbox{1.4 \AA}. Then, for each $R_{+}$, i.e. without
changing the parameters, the distance (see eq.~\ref{eq:def_d}) between
the energies calculated with a grid size $L_{n}$ (playing the role of
$V_{1}$) and the ones calculated with $L_{n+1}$ (playing the role of
$V_{2}$) was measured. The results are depicted in
fig.~\ref{fig:d_vs_dime}.

Two conclusions may be drawn from these data. One one hand, as the
size of the grid $L$ increases, the distance $d$ dimishes. This is
consistent with the expectation that, when the accuracy of an
approximation augments, the differences between an {\it exact}
potential energy and its approximated counterpart tend to
disappear. On the other hand, one sees that, for values of $L$ between
97 and 129, the algorithm implemented in APBS has practically
converged; in the sense that, for the worst case (namely, the one with
\mbox{$R_{+}=0.0$ \AA}), the distance between the energies calculated
with $L=97$ and $L=129$ is of the order of the thermal noise.

Two remarks must be made about this last fact. First, the situation
with \mbox{$R_{+}=0.0$ \AA} being the worst is easily understood if
one realizes that the discontinuity of the electric field in the
surface of separation is larger if the latter is closer to the
charges.  Thus, a greater sensitivity to details is expected in this
case.  Second, it must be stressed that the distances depicted in
fig.~\ref{fig:d_vs_dime} place a lower bound in the distances to be
considered meaningful when evaluating the robustness of the Poisson
energy.  For example, if the parameter $R_{W}$ is slightly changed
(keeping $R_{+}$ fixed to, say, \mbox{0.0 \AA}) any distance obtained,
using a grid size of 97 or 129, that is below $\sim RT$ (see
fig.~\ref{fig:d_vs_dime}) could not be associated to the lack of
robustness of the Poisson solvation energy in that particular region
of the parameter space, since it may be due to numerical inaccuracies
of the algorithm.

\begin{figure}[t]
\begin{center}
\epsfig{file=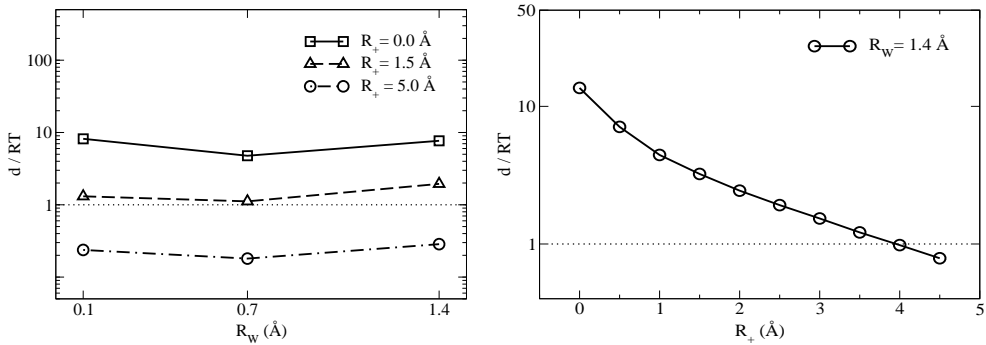,width=13cm}
\end{center}
\caption{Distance between instances of the Poisson solvation energy
corresponding to different values of the free parameters. The
\mbox{$y$--axis} corresponds to the distance $d$ measured in units of
$RT$ (with \mbox{$T=300$ K}) and the scale is logarithmic. Each point
represents the comparison of the energies calculated with the value of
the parameters in its $x$ coordinate and the one calculated with the
inmediately greater value (see text). The graphic on the left shows
the results obtained when $R_{+}$ is held fixed and $R_{W}$ is
varied. When $R_{W}$ is kept constant and $R_{+}$ is varied, the
measured distance $d$ is the one depicted in the graphic on the
right.}
\label{fig:d_vs_Rw}
\end{figure}

Having this in mind, let us fix the grid size to 97 or 129 depending
on the conformation\footnote{ What was actually done was to choose
$L=97$ or $L=129$ in order to keep the length of the grid cell below
\mbox{0.5 \AA} in each dimension. In practice, this led to using
$L=97$ for the most compact and globular conformations and $L=129$ for
the most extended ones.} and evaluate the sensitivity of Poisson
solvation energy to changes in the parameters $R_{W}$ and $R_{+}$ that
define the surface of separation, as an example of the application of
the distance $d$ to the second use proposed in
section~\ref{sec:introduction}. This is done in the particular region
of the parameter space which is typically explored in the literature:
for $R_{W}$, the values 0.1, 0.7, 1.4 (the Van der Waals radius of a
water molecule) and \mbox{2.8 \AA}; for $R_{+}$, the values from 0.0
to \mbox{5.0 \AA} in steps of \mbox{0.5 \AA}.  When $R_{+}$ is kept
constant and $R_{W}$ is varied, the results on the left part of
fig.~\ref{fig:d_vs_Rw} are obtained (only a few different values of
$R_{+}$ are depicted). When, in turn, $R_{W}$ is held fixed and
$R_{+}$ is varied, one obtains the results shown on the right side of
the same figure. In this second case, only the graphic corresponding
to \mbox{$R_{W}=1.4$ \AA} is depicted, as the results for different
values of $R_{W}$ are practically identical. As in
fig.~\ref{fig:d_vs_dime}, each point corresponds to the distance
between the instances of the Poisson energy with the $i$--th value of
the varying parameter and the one with the inmediately greater
$(i+1)$--th value.  Of course, if two instances with very distant
values of the parameters are compared, the measured distance is much
larger than the values depicted in fig.~\ref{fig:d_vs_Rw}. However,
this is not relevant to assess the robustness, since only the distance
between instances corresponding to slightly different parameters must
be small in order to render a fit significant.

From the data shown in fig.~\ref{fig:d_vs_Rw}, one may extract some
relevant conclusions. First, the two situations depicted are different
in an important sense: while the robustness increases ($d$ decreases)
as one moves towards larger values of $R_{+}$ holding $R_{W}$
constant, it does not change significantly in the opposite situation
(i.e. increasing $R_{W}$ with fixed $R_{+}$). The same behaviour may
be inferred from the fact that, on the left side of
fig.~\ref{fig:d_vs_Rw}, graphics corresponding to different values of
$R_{+}$ are found at different heigths, whereas, on the right side,
the data with different values of $R_{W}$ produce almost identical
results (this is not shown for the sake of visual comfort). The
second important fact that must be pointed out is that, in agreement
with what one would expect, the robustness of the Poisson solvation
energy is minimum when the surface of separation is placed close to
the molecule (i.e., small values of $R_{+}$).  In the left graphic of
fig.~\ref{fig:d_vs_Rw}, one sees that, when $R_{+}$ is of the size of
the water molecule radius (\mbox{1.4 \AA}), the distance between
instances of the potential energy produced by a small change in
$R_{W}$ approximately reaches the largest numerical indetermination in
fig.~\ref{fig:d_vs_dime} and, consequently, the Poisson energy may be
considered robust to such a change. In the right part of
fig.~\ref{fig:d_vs_Rw}, one finds that an equivalent level of
robustness is only achieved at values of $R_{+}$ around \mbox{3.0 \AA}
if $R_{W}$ is held fixed and what is changed is $R_{+}$.

To summarize, one may conclude that the robustness of the
\mbox{Poisson--based} electrostatic part of the solvation energy
steadily increases when the surface that separates the two dielectric
media is moved further away from the macromolecular solute. The
largest value of the distance $d$ is of the order of $10RT$ when the
surface of separation is placed on the Molecular or Van der Waals
Surface (\mbox{$R_{+}=0.0$ \AA}) and the sensitivity to parameter
changes approximately reaches the numerical indetermination of the
algorithm used when the surface is one layer of water molecules away
from the protein.

\section{Conclusions}
\label{sec:conclusions}

When calculating a term or the totality of a potential energy function
in complex systems, two situations are often faced: the necessity to
evaluate the relative accuracy of an approximation or an algorithm
respect to a more precise calculation and the need to assess the
significance of a free parameters' fit. Herein, a distance between two
different instances of the same potential energy function has been
devised, which may be used to answer the two preceding questions by
making meaningful statistical statements about the way in which energy
differences are modified when changing the algorithm or the
parameters.

In section~\ref{sec:application}, a practical example of the two cases
is given by studying the sensitivity of the \mbox{Poisson--based}
electrostatic part of the solvation energy to such changes. This
example is useful, on one hand, to show that the distance behaves
consistently in a real situation and, on the other hand, to estimate
the robustness of the Poisson energy when small changes are performed
on the ideal surface that separates the protein cavity from the
aqueous media. It is shown that this robustness, both to changes in
$R_{W}$ and in $R_{+}$, increases as the surface is moved further away
from the macromolecule, being \mbox{$d \sim 10RT$} when the surface is
placed at zero distance from the Van der Waals volume of the protein
and reaching the numerical indetermination at a distance of around a
layer of water molecules (\mbox{$\sim 3.0$ \AA}).

\begin{ack}
We would like to thank J.-L. Garc\'{\i}a Palacios, V. Laliena, and
A. Taranc\'on for illuminating discussions, Andr\'es Cruz for pointing
out an already corrected error in figures 1 and 2, and specially Prof.
N.A. Baker for providing valuable hints to compile APBS in our system.
This work has been partially supported through the following research
contracts: grant \mbox{BFM2003-08532-C02-01}, MCYT (Spain) grant
\mbox{FPA2001-1813}, Grupo \mbox{Consolidado} of the Arag\'on
Government ``Biocomputaci\'on y F\'{\i}sica de Sistemas Complejos''.
P. Echenique is supported by MECD (Spain) FPU grant.
\end{ack}

\end{document}